# It Ain't What You View, But The Way That You View It: documenting spreadsheets with Excelsior, semantic wikis, and literate programming

*Jocelyn Paine*
*www.j-paine.org/ and www.spreadsheet-factory.com/*
*popx@j-paine.org*

**ABSTRACT**

*I describe preliminary experiments in documenting Excelsior versions of spreadsheets using semantic wikis and literate programming. The objective is to create well-structured and comprehensive documentation, easy to use by those unfamiliar with the spreadsheets documented. I discuss why so much documentation is hard to use, and briefly explain semantic wikis and literate programming; although parts of the paper are Excelsior-specific, these sections may be of more general interest.*

## 1. INTRODUCTION

It is terribly important that computer programs should be easy for people to read and understand. Imagine that one of your clients needs an urgent fix to a program, but its author has gone on holiday or left the company; and although you have other staff with the general skills needed to fix it, none has ever worked on this particular program, or even on the phenomenon it models. How easily can they find its documentation, and how well will it tell them everything they need to know?

That, applied to spreadsheets, is the topic of this paper. I have been converting a 400,000-cell financial spreadsheet into code for my spreadsheet generator Excelsior [Paine, Tek and Williamson, 2006], in order that bespoke versions of the spreadsheet can quickly and reliably be produced for specific clients. Such a big spreadsheet needed extensive documentation. This led me to experiment with semantic wikis and with literate programming; the objective being to make the code the last thing that the modeller writes, ensuring that an easy-to-understand spec of each module has been written *first*. The experiments are preliminary, mainly concerned with implementation rather than evaluating readers' reactions, but may still interest others.

This paper is organised as follows. Section 2 is a general introduction to Excelsior. Section 3 diagnoses the diseases suffered by today's documentation, with reference to Parnas and Clements's paper *A rational design process: How and why to fake it* [Parnas and Clements, 1986]. Section 4 explains semantic wikis in general; Section 5 applies them to documenting Excelsior programs. Section 6 moves from wikis to the more general topic of literate programming — writing programs that are easy for people to read. Sections 7 and 8 are short evaluations and conclusions.

Some parts of the paper are Excelsior-specific, and will not apply directly to Excel. However, you may find the wiki material useful if you need to store off-spreadsheet documentation, or if you are using spreadsheet generators other than Excelsior, such as the one blogged in [Voyce, 2007]. The references on what's wrong with documentation, as well as on literate programming, I recommend to everyone.

## 2. BACKGROUND ON EXCELSIOR

At last year's EuSpRIG conference, I described how my co-author Emre Tek and I had converted a 10,000-cell housing-finance spreadsheet to Excelsior code, using Excelsior's features for listing spreadsheets as text, probing their structure, and "reverse-engineering" or "structure discovering" them. The paper is available as *Rapid Spreadsheet Reshaping with Excelsior: multiple drastic changes to content and layout are easy when you represent enough structure* [Paine, Tek and Williamson, 2006]. I shall summarise Excelsior briefly from that paper and show how I am using it in the current project.

Excelsior is a programming language for describing spreadsheets as sets of equations between groups of cells, or *tables*. The Excelsior *compiler* reads files containing the equations and generates an Excel spreadsheet. These files are submitted to it as plain text, accompanied by a *layout specification* indicating where the tables must go in the generated spreadsheet.

A single spreadsheet can be composed from more than one Excelsior file; indeed, Excelsior's most significant claim is that spreadsheets can be developed from modules which can be written, tested, and documented separately. This makes spreadsheeting very flexible: we can develop a whole family of spreadsheets, producing different but related versions merely by substituting one module for another, or changing parameters such as table sizes.

As well as a compiler, Excelsior has tools for listing existing spreadsheets and converting them *to* Excelsior code. This usually involves guessing which cells can be grouped together into tables, naming the tables, and relisting the spreadsheet to use these names rather than the original cell addresses. This *structure discovery* uses, amongst other things, a *run-detector* to find blocks of related formulae differing only by constant increments in a cell address or constant.

New since last year is a tool for displaying sheet-dependency networks. This works by collecting, for each sheet, references in its formulae to all other sheets. It outputs this information as statements in the graph-description language DOT used by the freeware Graphviz graph visualization package. I mention this because Graphviz is useful for displaying well-laid out network diagrams, and may help others needing to diagram spreadsheets. To those trying it, I also recommend the interactive ZGRViewer. See [Graphviz], [DOT] and [ZGRViewer].

### 2.1 Excelsior Code Example

This example is from Section 2 of [Paine, Tek and Williamson, 2006]. Consider a spreadsheet for forecasting housing numbers. Let's assume we have the one-dimensional tables `Builds` and `Demolitions` that give the number of houses built and demolished per year, and `Net` that gives the net new houses per year. Then, if years 2000 and 2001 are valid subscripts for these tables, we could write:

```
Net[ 2000 ] = Builds[ 2000 ] - Demolitions[ 2000 ]
Net[ 2001 ] = Builds[ 2001 ] - Demolitions[ 2001 ]
```

Tables can have more than one dimension. We could make the type of house a second dimension:

```
Net[ 2000, 1 ] = Builds[ 2000, 1 ] - Demolitions[ 2000, 1 ]
```

Excelsior has an "all" construction which generates a set of equations relating all elements of tables. Thus, Excelsior will expand the following into as many equations as there are elements of `Builds`, `Demolitions` and `Net`:

```
Net[ all y, all ht ] = Builds[ y, ht ] - Demolitions[ y, ht]
```

## 2.2 Listing And Run-detecting Spreadsheets

Excelsior can list the formulae in an existing spreadsheet, for example:

```
'House Stocks'!F1 = "Newly built houses"
'House Stocks'!G1 = "Demolished houses"
'House Stocks'!H4 = 'House Stocks'!F4 - 'House Stocks'!G4
```

This can be useful, but if one formula is repeated many times, the listing will become too large and repetitious to be intelligible.

To avoid this, we can run-detect a sheet and *then* list it. When listing run-detected sheets, Excelsior uses a special notation to indicate repetition. For example,

```
'House Stocks'!H[V0 in 4:13] =
  'House Stocks'!F[V0] - 'House Stocks'!G[V0]
```

Here, the `[V0 in 4:13]` indicates repetition down a column. `V0` is a row variable; in the listing it varies over rows 4 to 13. The equation is therefore short for:

```
'House Stocks'!H4 = 'House Stocks'!F4 - 'House Stocks'!G4
...
'House Stocks'!H13 = 'House Stocks'!F13-'House Stocks'!G13
```

As another example, suppose listing with run-detection displayed this equation:

```
'House Stocks'![C:D]1 = "Year"
```

This uses the same notation, but between columns rather than rows. The equation means that cells C1 and D1 of the sheet both contain `"Year"`.

Finally, suppose that run-detection and listing displayed this:

```
'Home Sales'![V0 in C:D][V1 in 4:13] =
   'House Sales'![V0+2][V1-1] - 'Flat Sales'![V0+3][V1+1]
```

This is short for:

```
'Home Sales'!C4 = 'House Sales'!E3 - 'Flat Sales'!F5
...
'Home Sales'!D13 = 'House Sales'!F12 - 'Flat Sales'!G14
```

### 2.3 Run Detection With Renaming

Consider again the equation:

```
'House Stocks'!H[V0 in 4:13] =
   'House Stocks'!F[V0] - 'House Stocks'!G[V0]
```

It seems likely that cells H4:H13, F4:F13, and G4:G13 form three distinct tables. We could tell Excelsior to relist the spreadsheet while giving H4:H13 the name `Net`, F4:F13 the name `Builds`, and G4:G13 the name `Demolitions`. If further, we told it that row 4 corresponds to subscript 2000, and column F to subscript 1, listing would give the equation:

```
Net[ all V0, all V1 ] =
   Builds[ V0, V1 ] - Demolitions[ V0, V1 ]
```

These are clearly equivalent to the first equation, after renaming the cell groups as stated. This is the essence of how I have been converting the 400,000-cell spreadsheet mentioned in my introduction. Even if one doesn't want to convert a spreadsheet to Excelsior, listing it with run-detection and renaming is a very good way to make it intelligible.

This, therefore, is how one can convert a spreadsheet to Excelsior code. But code is never enough. Code can only tell you how a program does behave, not how it should behave**.** This brings me to an important question.

### 3. WHY IS DOCUMENTATION SO BAD?

The rest of this paper describes approaches to documenting Excelsior programs so they are easy for people to read. I shall start by referring to a paper that makes valuable points about why documentation is bad and how to improve it, and is relevant to my work with wikis. I thoroughly recommend it to all programmers.

### 3.1 We Must Fake Rationality

This paper is *A Rational Design Process: How and Why to Fake It* by David Parnas and Paul Clements [Parnas and Clements, 1986]; [Bredereke, 2002] is a summary. The authors argue that although we aspire to be rational software designers, we don't act rationally. We may, for example, use a technique because we like experimenting with it, rather than because it's best for our project.

Nevertheless, even though we can't follow a rational design process completely, we should do so as closely as we can. Moreover, we should write the documentation we would have produced if we had followed the ideal process. We should "fake a rational design process".

### 3.2 Why Is Documentation Hard To Use And Infrequently Read?

Section VI.A of the paper diagnoses the problems afflicting documentation. (It was published 20 years ago, but the problems remain). The authors state these to be (I quote):

> (1) Poor organisation. Most documentation today can be characterised as "stream of consciousness," and "stream of execution." "Stream of consciousness" writing puts information at the point in the text that the author was writing when the thought occurred to him. "Stream of execution" writing describes the system in the order that things will happen when it runs. The problem with both of these documentation styles is that subsequent readers cannot find the information that they seek. It will therefore not be easy to determine that facts are missing, or to correct them when they are wrong. It will not be easy to find all the parts of the document that should be changed when the software is changed. The documentation will be expensive to maintain and, in most cases, will not be maintained.

> (2) Boring prose. Lots of words are used to say what could be said by a single programming language statement, a formula or a diagram. Certain facts are repeated in many different sections. This increases the cost of the documentation and its maintenance. More importantly, it leads to inattentive reading and undiscovered errors.

> (3) Confusing and inconsistent terminology. Any complex system requires the invention and definition of new terminology. Without it the documentation would be far too long. However, the writers of software documentation often fail to provide precise definitions for the terms that they use. As a result, there are many terms used for the same concept and many similar but distinct concepts described by the same term.

> (4) Myopia. Documentation that is written when the project is nearing completion is written by people who have lived with the system for so long that they take the major decisions for granted. They document the small details that they think they will forget. Unfortunately, the result is a document useful to people who know the system well but impenetrable for newcomers.

On a personal note, I notice such problems in my own software documentation. Suppose I have written one module that defines a data structure and several others that use it (in any programming language, not just Excelsior). I am often tempted to explain in each module those facts about the structure that are most relevant. Thus, I may end up with several explanations of the structure. Each overlaps the others in coverage, without being complete in itself. Later, I may rename a subsidiary structure or an operation in some modules but not others. Inconsistencies multiply, and a reader must examine the comments in many different modules, trying to reconcile them.

## 4. SEMANTIC WIKIS

The problems noted above include duplication of facts, difficulty in finding information, and lack of consistent and precise terminology, leading to ambiguity and confusion in naming things. These led me to experiment with semantic wikis.

A semantic wiki is an interlinked collection of Web pages which can be collaboratively edited through a Web browser, and which can automatically generate (amongst much other content) indices showing which pages link to which other pages. By convention, each page is written to be about one topic. Topics can be given a category, and the indexing tools can subdivide indices by category.

This seemed a promising framework for Excelsior documentation, because we could write one page for each module or important concept. Links between pages would indicate, for example, which modules use which other modules: thus the wikis's organisation would reflect the spreadsheet's structure. This structure would clarify where particular topics should be documented, making it easier to find information and eliminating duplication. Ambiguity and confusion in naming would be reduced because it is a wiki convention to use a page's title as the name of its topic. For a link to work properly, the person who writes it must therefore use its correct name.

I shall now explain how wikis can be edited, displayed, and implemented. In the section following this one, I shall explain how I applied this to Excelsior.

### 4.1 Wikis

A wiki, is, as stated above, a Web site that allows collaborative editing. The word has become associated with a particular style of notation for writing the pages, and a particular way of displaying them. There are many different implementations of wikis, and many wiki sites: the best known site is Wikipedia [Wikipedia].

### 4.2 Wiki Markup

Because wikis are Web sites, their pages ultimately get converted to HTML. One could, therefore, write them in HTML. However, HTML's tags and brackets are distracting, making it hard to focus on content. So although some wikis do let you insert HTML, they also have their own simple markup language. The "wiki engine", i.e. the program that runs on a Web server and handles tasks such as editing, deleting, and creating pages, will render the markup into HTML before sending it out to a browser.

Here is an example, from the Wikipedia article about Yodel Bank:

```
'''Yodel Bank''' was an online [[anonymous banking]] system
which ended operations during November of 2005. Yodel Bank
```

```
was not a registered company in any country, its operator's
identity is  unknown, and it existed entirely outside any
countries' laws.

==Operations==

Yodel Banks operations functioned on top of various forms of
[[electronic money]] including the [[Digital Monetary
Trust]], [[e-gold]], [[Pecunix]], [[1MDC]],
[[FreeTraders]], and the e-[[Liberty Dollar]].
```

The three quotes surrounding the first two words display them in bold. Double square brackets around a term mark it as a link to another article. The equals signs indicate that `Operations` is a section heading.

### 4.3 Semantic Wikis

A semantic wik is a wiki crossed with a database. Pages can include *semantic markup*, also known as *semantic annotations*. Semantic markup can assert facts, about which the wiki's users can ask questions using a *semantic search* interface. The wiki engine will generate a page on the fly to answer the search.

A page's semantic markup can contain questions as well as facts. The wiki will again answer them, but this time, will render the answer as part of the page. This is how one includes machine-generated, but human-readable, content in a page. For example, suppose we have a semantic wiki with a page for each capital city. Suppose also that each of these pages has semantic markup stating which country its city is capital of. Then we could write a page about capitals in general, and place in it a question asking for all the capitals to be listed with their countries. Whenever anyone visited that page, the wiki would generate such a list; so it would always be up to date, no matter how many capital-city pages were added or deleted.

### 4.4 Links Are Facts; Indices Answer Questions About What Links Where

Semantic wikis typically treat links as facts. A link from page A to page B is treated as the fact "Page A links to page B". If another page's semantic markup includes a question that asks "which pages link to page B?", the wiki will answer it by, in effect, generating an index of links. Typically, the person writing the markup can indicate how the index should be laid out, for example as a list of comma-separated words, or as a bulleted list.

### 4.5 Implementing A Semantic Wiki

Anyone can mount a wiki on their Web server. There are many different wiki programs available; I used the same one that Wikipedia does, MediaWiki [MediaWiki]. This was partly because it is free and well-tested, with many experienced users and good support. Also, I wanted to run it on my Web site. This is a commercial internet service provider (Mythic Beasts Ltd.), and the facilities provided by ISPs are often restricted, so the wiki had to work with what I had available. As it happens, Mythic Beasts ran MediaWiki for their own work, liked it, and recommended I install my own copy.

Installing MediaWiki involved downloading the distribution and then following the installation instructions. This was straightfoward, but it helps to have experience with software installation and Web servers, so that you can recognise and correct typos and other blunders. Otherwise, it is worth asking an expert.

MediaWiki can be converted to a semantic wiki by installing Semantic MediaWiki [Semantic MediaWiki] on top of it. This also was straightforward.

## 5. ORGANISING THE WIKI: MODEL PAGES, MODULE PAGES, CONCEPT PAGES, AND INDICES

I have explained how one writes wiki content, how semantic wikis generate content from semantic markup, and that one can run a wiki on one's own Web site. But how did I structure the knowledge within a semantic wiki so as to document an Excelsior-generated spreadsheet? That is the subject of this section.

### 5.1 Model Pages And Module Pages

The Excelsior version of any reasonably interesting spreadsheet will contain more than one module. (For the purposes of this paper, modules are the separate source files that together constitute the Excelsior code of a complete spreadsheet.) Our documentation must therefore distinguish between a spreadsheet and its component modules.

In a wiki, as Section 4 explains, it is conventional to have one page per topic. Semantic wikis allow each topic to be given a category. (Strictly speaking, some non-semantic wikis do too, but with far less flexibility in using them.) To apply this, I decided to have two categories of page: *model pages* and *module pages*. (To avoid confusing mathematical readers, I should say that I use the word "model" in the informal sense in which many people talk about "spreadsheet models", not as in "model theory". I use the word "category" in the sense of "classification", not as in "category theory".)

A module page documents a module. A model page documents an entire spreadsheet, and should at least state the modules from which the spreadsheet is composed. It can do so by linking to their pages: the wiki's author can then write semantic markup to generate, for each module, a list of the models that use it.

As an example, the following question, placed within a page, will list all the pages that have that page as a module. For an explanation, see [Help:Inline queries]:

```
<ask>[[has module::{{PAGENAME}}]]</ask>
```

### 5.2 Concept Pages

Section 3.2 discussed the problems that arise when there is no clear place to put information. To avoid this in the wiki, I added another category of page, the *concept page*. The idea is that any information relevant to more than one module should live in its own concept page, to which other pages will link if they want to refer to the information. The page's title, which must be unique within the wiki, can be an unambiguous name for the concept.

Although I've not evaluated its merits, I would expect a good writing style for concept pages to be the *inverted pyramid* [Inverted pyramid], favoured by journalists who must convey key facts quickly to a probably bored and inattentive reader. This style presents the gist immediately and succinctly, before descending into gradually increasing levels of detail. Hopefully, staff new to the material, who may have to read it at short notice, can thus assess its relevance before being overwhelmed by detail.

**5.3 Semantic Wikis As An Interface To Excelsior**

This section is more technical than the rest, but may interest wiki specialists. Its point is that the list of modules on a model page (Section 5.1) is an essential part of the Excelsior program defining the spreadsheet. If we're going to include it in the wiki, shouldn't we also include the rest of the program?

This leads to the idea of using the wiki as an interface to Excelsior; that we could write a list of modules and then submit it to a wiki add-on that finds the modules it links to, extracts their code, invokes Excelsior to compile it, and delivers up the resulting spreadsheet over the Web. Something similar might also be done with other spreadsheet generators than Excelsior.

Such add-ons are feasible, but I haven't experimented with them. A decision needed here is how to represent the Excelsior code. The information about which modules constitute a spreadsheet is on the page as links, which the semantic wiki regards as semantic annotations (Section 4.4). Should the Excelsior code also be semantic annotations? This could make it possible for the wiki to generate table cross-reference listings, for example; or even cross-reference graphs, which could be generated by the nice graphing software written by "Tels" [Tels]. However, it would require extra markup, making the code harder to read and write.

It also hits one specific problem. Semantic MediaWiki allows one to write *binary relations* in the semantic annotations. A binary relation (this is standard mathematical terminology) relates two items, as in the statements "1 is less than 2", "Model A uses module B", or "Page A links to page B". However, Semantic MediaWiki does not allow one to relate an arbitrary number of items, and this might be needed. There is a heavily modified version that does — used in BOWiki [BOWiki] — but it is not included with the standard distribution.

**6. LITERATE PROGRAMMING**

As an alternative to the wiki, I have been experimenting with "Literate Excelsior", a version of Excelsior where the syntax of source files is modified to make it easier to document the code in them. This allows me closer control over the syntax than does the wiki, and also makes it possible to work away from an internet connection.

*Literate programming* is the discipline of writing programs so that they're easy for people to read. This name, well-known in computing, was coined by Donald Knuth, who advertises his book [Knuth, 1992] on the topic by saying [Knuth, b]:

> Literate programming is a methodology that combines a programming language with a documentation language, thereby making programs more robust, more portable, more easily maintained, and arguably more fun to write than programs that are written only in a high-level language. The main idea is to treat a program as a piece of literature, addressed to human beings rather than to a computer. The program is also viewed as a hypertext document, rather like the World Wide Web. (Indeed, I used the word WEB for this purpose long before CERN grabbed it!)

Chris Lee gives a good summary at [Lee, 2000]. A nice collection of examples can be found at [Literate Programming Wiki] which calls each of its 407 articles "simultaneously a document and a piece of code that you can view, download, compile, and run".

## 6.1 Input To Literate Programming Systems

In most literate programming systems, documentation goes in the same input file as the code it documents. Such systems therefore contain both a compiler and a tool that generates the display form (usually HTML) of the document. The literate programming system must therefore be able to distinguish code from documentation, and may impose conventions to do so. Thus the literate-programming form of a programming language is usually different from its standard form.

## 6.2 Literate Excelsior

Following the above, whereas "plain" Excelsior uses comment markers to delimit non-code text, I have implemented Literate Excelsior so it treats all text in input files as comment, unless indented by at least two spaces, when it becomes code. I found this an easy convention to follow when editing, and it reduces clutter both when editing and when reading the final documentation. A preprocessor extracts the code and feeds it to Excelsior's compiler.

A separate preprocessor converts Literate Excelsior files to Web pages for display as documentation. This styles code as Wikipedia does, indenting it and setting it off within a pale blue box, thus making it immediately clear which parts of the document are code, without being distracting. The preprocessor recognises some simple wiki-style markup, such as equals signs to indicate a heading. I implemented this for the reasons noted in Section 4.2: such markup is easier to write than is HTML.

## 6.3 Narrative Style And The Order Of Program Pieces

The order in which things are explained is important for ease of understanding. Thus, a literate programming system must let you arrange your text in the order you deem most intelligible. Since code should be near the text that explains it, the system must let you do the same with the source code.

Many programming languages don't allow such freedom. Pascal, for example, has rigid rules about the order of declarations. This is not a problem with Literate Excelsior, because the equations making up a program can be written in any order.

However, text ordering may be one area where literate programming will cause problems. People use documentation for many purposes. An order that works when reading about a program for the first time may not help once you know the program and need to find something specific within it. (I find this a problem with language books: a chapter that introduces some easy touristic conversation in Portuguese or Greek is usually not a good reference.)

**6.4 Literate Excelsior And Function Introduction**

A very common pattern in Excel is the following conditional:

```
IF( NOT(ISNA(expr)), expr, other_expr )
```

This is testing whether `expr` yields a sensible result, and returning it if so. If not, `other_expr` is returned. Similar patterns are seen, for example, in testing whether an expression is zero and dividing by it if not.

When `expr` is complicated, such formulae are hard to read. To mitigate this, Excelsior allows the programmer to define *macros*: functions which it will *macro-expand* (replace by their definitions) before feeding the formulae to Excel. Thus the programmer need only write `expr` once. Macros can be defined anywhere in a source file. This gives extra freedom when writing literate programs, because one can code important parts of a formula as macros, thus making them a named item in their own right, which can have its own explanation. The macro, and its explanation, can go either before or after the formula that calls it: sometimes one seems appropriate, sometimes the other.

**7. EVALUATION**

The work described here is preliminary, and I am still its main user, so there is not yet much to evaluate. Installing Semantic MediaWiki is straightforward, even under a commercial internet service provider. It is easy to define different categories of page within it, and to automatically generate link indices. The precise naming and lack of duplication enforced by concept pages is useful. However, when writing them, it was sometimes difficult to decide, when describing a concept that could be categorised into subconcepts, whether to give subconcepts their own page. Similarly, when concept pages already written turned out to be closely related, it was difficult to decide when to merge them. I would be interested to hear of others' experiences.

As with the wiki, I am still the main user of Literate Excelsior. As explained in Section 6.2, the change in notation from "plain" Excelsior is small: no explicit comment markers, all text being documentation unless indented. But this makes a surprisingly big change to the "feel" of the writing: much more like writing expository text than like coding. Perhaps this will fade; at the moment though, it feels like a good thing.

**8. CONCLUSION**

One thing seems clear. Programmers — including spreadsheeters — need to document. They must therefore be good expository writers. A good starting point is Michael Covington's *How to Write More Clearly, Think More Clearly, and Learn Complex Material More Easily* [Covington, 2002]. For examples of clear writing, I recommend any volume of essays by Isaac Asimov.